\documentclass[aip,reprint,graphicx]{revtex4-1}

\usepackage{graphicx}
\usepackage{dcolumn}
\usepackage{bm}

\usepackage[utf8]{inputenc}
\usepackage[T1]{fontenc}
\usepackage{mathptmx}

\usepackage{multirow}
\usepackage{listings}
\usepackage{subfigure}

\usepackage{color}
\usepackage{seqsplit}

\usepackage[pdftex,
    bookmarks,
    bookmarksopen=true,
    bookmarksnumbered=true,
    pdfauthor={C.~A.~Knapek, D.~P.~Mohr, P.~Huber},
    pdftitle={Pulsed Complex Plasma in Microgravity},
    pdfkeywords={complex plasma, microgravity, void, void-free, pulsed plasma, void-closure, radio-frequency discharge, parabolic flight},
    colorlinks, linkcolor=blue, urlcolor=blue
]{hyperref}

\makeatother

\usepackage{academicons}
\usepackage{xcolor}
\newcommand{\orcid}[1]{\href{https://orcid.org/#1}{\includegraphics[width=3mm]{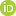}}}

\begin{document}

\title[Pulsed Complex Plasma in Microgravity]{Pulsed Complex Plasma in Microgravity}

\author{C.~A. Knapek \orcid{0000-0001-7105-627X}}
\email[]{christina.knapek@physik.uni-greifswald.de}
\affiliation{Institute of Physics, University of Greifswald, 17489 Greifswald, Germany}
\affiliation{Institut f{\"u}r Materialphysik im Weltraum, Deutsches Zentrum f{\"u}r Luft- und Raumfahrt, 51147 K\"oln, Germany}

\author{D.~P.~Mohr \orcid{0000-0002-9382-6586}}
\affiliation{Institute of Physics, University of Greifswald, 17489 Greifswald, Germany}
\affiliation{Institut f{\"u}r Materialphysik im Weltraum, Deutsches Zentrum f{\"u}r Luft- und Raumfahrt, 51147 K\"oln, Germany}

\author{P.~Huber \orcid{0000-0002-1426-6104}}
\affiliation{Institut f{\"u}r Materialphysik im Weltraum, Deutsches Zentrum f{\"u}r Luft- und Raumfahrt, 51147 K\"oln, Germany}

\date{\today}

\begin{abstract}
A new experimental method for creating void-free complex (``dusty'') plasmas under microgravity conditions is presented. The method is based on a pulsed operation mode of a four-channel radio-frequency generator for plasma sustainment.
A dust cloud of micrometer-sized particles can be immersed in the bulk of a low temperature plasma under microgravity conditions. It typically contains a central volume depleted of particles -- the void -- that prevents the generation of large, continuous clouds.
Experiments performed at different neutral gas pressures and discharge volumes during the microgravity phase of a parabolic flight show that the central void is closed completely once the pulsed operation mode is applied.
The particle cloud shape, and the density distribution within the cloud, are practially independent on the pulse period within the investigated parameter range, and mainly depend on the overall discharge parameters neutral gas pressure and discharge volume.
This indicates that the pulsed operation of the plasma source does not introduce new physical effects on the particles aside from the void closure.
The proposed method has great potential for future application in experimental facilities dedicated to fundamental studies of large three-dimensional, homogeneous complex plasma systems in microgravity.
\end{abstract}


\maketitle

\section{\label{sec:introduction}Introduction}
        
 Laboratory complex plasma are low-temperature plasmas containing nano- to micrometer sized particles as an additional component \cite{morfill2009,fortov2005,piel2016,melzer2019}. In the plasma, the particles acquire negative charges up to several thousands of elementary charges, with the charge mainly depending on the particle size and the electron temperature $T_{e}$ of the plasma. The particles interact via a screened Coulomb potential, and form strongly coupled many-particles systems in a variety of states (solid, liquid, gaseous).

  The particles are individually observable by optical diagnostics, and fundamental processes of many-particle physics can be studied, e.g. phase transitions \cite{rubin-zuzic2006,knapek2007}, nonlinear waves \cite{sun2018,tsai2016,kananovich2020}, or turbulence \cite{schwabe2017,lin2020}.
 
  Gravity forces the particles to sediment into the plasma sheath region, where strong electric fields balance the gravitational force. The resulting systems are vertically compressed and limited by the extent of the sheath. For the investigation of large homogeneous three-dimensional (3D) clouds, an environment with considerably reduced gravity is needed, e.g. a facility on the International Space Station (ISS) \cite{nefedov2003,thomas2008,pustylnik2016}, or parabolic flights. Once the gravitational force is removed, the weaker electric forces in the bulk plasma are sufficient to confine particles to the plasma bulk region.

\begin{figure}
\includegraphics[width=\columnwidth]{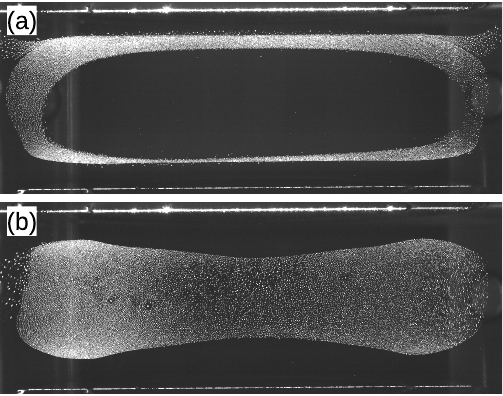}
\caption{\label{fig:fig1} The images show cross-sections through particle systems created in microgravity during a parabolic flight. The particles scatter incident laser light used for illumination (details of the experiment procedure are given in section~\ref{sec:setup}). The experimental parameters are: electrode separation 45~mm, neutral gas pressure 20~Pa, peak-to-peak voltage at the electrode segments 70~V. The particles are melamine-formaldehyde spheres with a diameter of  4.41$\mu$m. (a) Large central void in a continuous radio-frequency discharge. (b) No void in a pulsed radio-frequency discharge (pulse base period: 200$\mu$s, see section~\ref{sec:experiments} for details).}
\end{figure}

In a typical electropositive radio-frequency (rf) discharge of a noble gas, the electric field that causes the inwards pointing confining force on the particles in turn accelerates the positive ions outwards. The ion flow excerts an outward-pointing ion drag force on the particles \cite{khrapak2005,hutchinson2006}.
This often results in the formation of the so-called void -- a region in the discharge center which is depleted of particles \cite{nefedov2003,morfill1999,goree1999}. A typical example of a void is shown in Fig.~\ref{fig:fig1}(a).

While the physics behind void-formation can be subject of investigations in itself \cite{pustylnik2017}, large homogeneous three-dimensional clouds without the central void are favored to study e.g. statistical properties of many-particle systems.

It has been shown that the void can be successfully closed by reducing the discharge power, and thus the ion drag force \cite{lipaev2007}. One drawback of this method is that the discharge is operated at powers close to the minimum power needed to maintain it. It can easily terminate, especially in the presence of particles drawing electrons from the discharge onto their their surface.

An alternative method to manipulate discharge parameters is the plasma generation by a pulsed power source.
A plasma sustained by a pulsed source is in the afterglow during the off-times, i.e. the afterglow is temporal, and it is re-ignited again during the on-time. Global models that describe the evolution of the pulsed plasma can be found in the literature \cite{lieberman1996,lieberman2005, liu_fx_2015}. Pulsed discharges were for example studied as a tool to modify the gas phase chemistry and manipulate ion bombarding energies in the scope of industrial plasma processing \cite{Booth_1997}, or to control the electron energy distribution function \cite{Song_2012}. 

 During the temporal afterglow, electrons and ions diffuse towards the walls. Dust particles previously charged in the plasma by continuous streams of electrons and ions to their surface loose some of their charge until the charge becomes frozen \cite{ivlev2003}.
Since there is no production of electrons and ions, and the electrons are more mobile, the diffusion changes from ambipolar to free, and the remaining charge on dust particles in the discharge is determined by the nature of this transition \cite{couedel2008,couedel2022}. 
The particles themselves can in turn influence the plasma parameters of the decaying plasma by e.g. emitting previously collected electrons from their surface due to collisions with metastables \cite{denysenko2021,denysenko2022}. Even positive charges have been measured on particles in temporal afterglows \cite{chaubey2021,chaubey2022}, and the effect of the cooling of the background gas in the former plasma region during the off-time on the neutral gas drag experienced by the particles has been studied in a spatio-temporal afterglow \cite{vanhuijstee2022}.
Spatial afterglow, where the afterglow is located in a different region than the region of active plasma production, has further been used to investigate particle decharging \cite{vanMinderhout2021}.

Here, we will present a new method for void-closure utilizing a four-channel pulsed rf source, and investigate the effects on the particle cloud. 
The method has the advantage that the average rf power input into the plasma, and thus the plasma density, can be kept constant, while pulse parameters are varied.
A series of experiments performed in microgravity during parabolic flights at different neutral gas pressures, discharge volumes and pulse periods demonstrates the effectiveness of the method, even over a wider parameter range.
Some basic properties of this  ``pulsed complex plasma'', such as the spatial particle distribution, and the homogeneity of the particle density are analyzed, and their dependence on the experimental parameters is examined.

The paper is organized as follows. The experimental setup is described in Section~\ref{sec:setup}, the experimental procedure and specific parameters for the performed experiments are given in Section~\ref{sec:experiments}. The data analysis methods and respective results are presented in Section~\ref{sec:results}, followed by a discussion of the possible underlying mechanisms for the observed behaviour in Section~\ref{sec:discussion}.
The paper is concluded in Section~\ref{sec:conclusion}, with an additional outlook for future works.
 
\section{\label{sec:setup}Experimental Setup}

\begin{figure}
\includegraphics[]{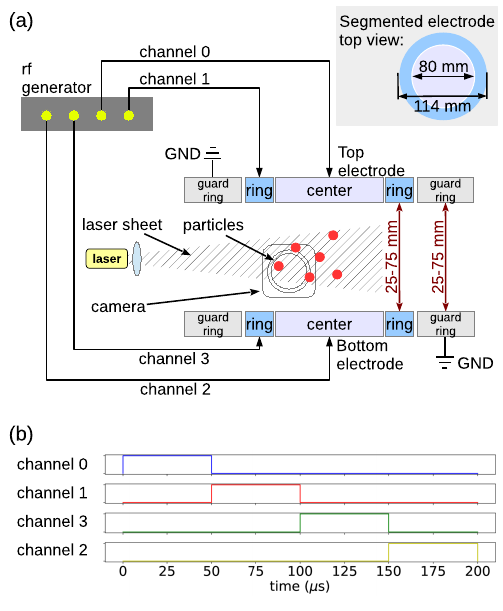}
\caption{\label{fig:fig2} (a) Schematic of the experimental setup with two segmented electrodes mounted parallel to each other inside the plasma chamber. Each segmented electrode consists of a central disk electrode (light blue) and a ring electrode (dark blue), as illustrated in the inset showing the top view. The electrodes are surrounded by grounded guard rings (gray). Each of the segments is connected to an individually adjustable channel of a four-channel rf generator. Experiments are performed in the following way: A plasma is ignited between the electrodes, micrometer sized particles are injected and a cross section of the particle system is illuminated by a vertically spread laser sheet. The scattered light is then recorded by cameras. (b) Example of the operation of the four rf channels in a sequential pulsed mode. The on-time of each channel is the time when the shown signals are high. Here, each of the channels is pulsed with a base period of $50$~$\mu$s, a period of $200$~$\mu$s, and an offset between the channels chosen such that only one channel is on at a time. Repeated infinitely, this type of pulsing would lead to a ``circular'' switching pattern between the electrodes: top center $\rightarrow$ top ring $\rightarrow$ bottom ring $\rightarrow$ bottom center.}
\end{figure}

Experiments were performed in the Zyflex plasma chamber \cite{knapek2021}. The chamber is a cylindrical Aluminium vessel with outer diameter of $270$~mm and a height of $250$~mm. Two segmented electrodes are mounted parallel to each other in the chamber, each consisting of two electrically isolated parts: one central disk-shaped electrode with diameter of $80$~mm, surrounded by one ring-shaped electrode with outer diameter of $114$~mm as shown in Fig.~\ref{fig:fig2}(a). 
The separation between the segmented electrodes can be adjusted in the range $25-75$~mm by piezo motors mounted inside the chamber. For a detailed technical description of the chamber see \cite{knapek2021}.

Each of the electrode segments is connected to its own rf channel of a four-channel radio-frequency (rf) generator operating at $13.56$~MHz. The amplitude of each channel, as well as the phases of signals between the channels, can be individually adjusted, while all channels share the same time base. Each channel can be either operated in a continuous mode, or in a pulsed mode as described in detail below.

After ignition of an Argon plasma by applying rf signals to the electrode segments, micrometer-sized particles are injected into the plasma from dispensers mounted in the chamber side walls.
The particles are illuminated by a $660$~nm diode laser with $60$~mW maximum optical output power and beam shaping optics. The optics (Sch\"after+Kirchhoff macro line generator) produce a vertically spread, horizontally thin, laser sheet with uniform intensity profile in line direction and an extended range of focus of $82$~mm. The laser is mounted such that it illuminates the central vertical cross-section of the chamber. The scattered light is recorded by a digital camera with an 8.9~MPixel sensor at a frame rate of 70 frames per second and a spatial resolution of 40~$\mu$m/pixel.

For operating in the pulsed mode an Arduino Nano microcontroller provides trigger signals for each rf channel. The trigger signals are determined by a pre-defined pulse sequence supplied by the user, and uploaded to the microcontroller.

A sequence can contain up to $1024$ entries (limited by the memory of the microcontroller), describing the on/off states of the four channels at consecutive time steps of the length $T_{\rm{base}}$ (the base period).
The sequence is repeated in an infinite loop.
Periodic pulsing, for example, can be achieved by combining short pulses, as illustrated in Fig.~\ref{fig:fig2}(b): all four channels are pulsed with a base period of $T_{\rm{base}}=50$~$\mu$s, a period of $4T_{\rm{base}}=200$~$\mu$s, a pulse width of $1T_{\rm{base}}=50$~$\mu$s, with an offset of $0$, $50$, $100$ and $150$~$\mu$s for channels $0,1,2,3$, respectively. In that example, at each time one channel is on.

Alternatively, a long random sequence can be repeated in a loop to realize e. g. pseudorandom pulsing of the four channels.

\section{\label{sec:experiments}Experiments}

\begin{table}
 \caption{\label{table:experiments} Experimental parameters for the data sets \#1-15: Electrode separation $\Delta L$, neutral gas pressure $P$, peak-to-peak voltage $V_{\rm{pp}}$ and pulse base period $T_{\rm{base}}$.}
  \begin{ruledtabular}
     \begin{tabular}{ccccc}

     data set \# & $\Delta L$ [mm] &$P$ [Pa]  & $V_{\rm{pp}}$ [V] & $T_{\rm{base}}$  [$\mu$s]\\\hline\hline
   
    1 &75 &20 &70 &50 \\\hline
    2 &75 &20 &70 &50 \\\hline
    3 &75 &20 &70 &100 \\\hline

    4 &75 &5 &70 &50 \\\hline
    5 &75 &5 &70 &50 \\\hline
    6 &75 &5 &70 &100 \\\hline
    
    7 &45 &20 &70 &50 \\\hline

    8 &45 &20 &70 &200 \\\hline
    
    9 &45 &20 &70 &400\\\hline

    10 &45 &5 &70 &50 \\\hline
    
    11 &45 &5 &70 &50 \\\hline

    12 &45 &5 &70 &100 \\\hline

    13 &45 &5 &70 &800 \\\hline
       
    14 &45 &5 &90 &50 \\\hline
    
    15 &45 &5 &90 &400\\
 
  \end{tabular}
  \end{ruledtabular}
\end{table}

The experimental setup is integrated in racks that can be mounted into an airplane that performs a special flight maneuver (parabola). During a parabola, the forces on the microparticles due to gravity are reduced down to the order of $10^{-2}$g for $\approx 22$~s. Typically, $31$ parabolas are flown during one parabolic flight campaign day.

The experiments were conducted as follows: before each parabola, electrode separation, neutral gas pressure, and the voltages applied to the electrodes were adjusted to desired values. Pulsing of the plasma was obtained by consecutive on/off-switching of the four rf channels in a pseudorandom, pre-calculated sequence.
The sequence determines, for each time step in units of the base period $T_{\rm{base}}$, one channel to be in the on-state, while all others are switched off. This ensures that at each time, only one rf channel is used to drive the plasma, and that there is no short-term periodic pattern in the succession of channels. 
Thus, due to the geometric arrangement of the four electrodes, the location of plasma generation is continuously shifted between center (top or bottom) and ring (top or bottom) electrodes in a random fashion. Since one channel is always on, strictly speaking the pulsed plasma does not reach a temporal afterglow state, but rather a spatial and temporal shift of the plasma production region. Therefore, the underlying physics might differ from those in real temporal afterglow. In temporal afterglow, electrons and ions diffuse outwards from the former plasma region. In the presented configuration though, an new plasma region is generated immediately after switching off the source of the initial region, and the diffusion of ions and electrons that were generated in the initial region would be under the influence of the (adjacent) region of ``newly'' generated plasma.

The exact sequence utilized for the experiments is given in the Appendix~\ref{sec:appendixA} with additional details on its generation and properties. The only free parameter is $T_{\rm{base}}$, which is the minimum time a channel can be in the on-state. Since the length of the sequence is limited by the available memory of the microcontroller, it is continously repeated from the beginning after completion. This can in principle introduce some kind of long-term periodicity.

At the beginning of each parabola, microparticles (spherical melamine-formaldeyde particles with a diameter of $4.41$~$\mu$m) were injected into the plasma and images of the particle distribution were recorded during the $22$~s of reduced gravity.

The injection of particles is implemented via the principle of a salt shaker: a reservoir with particles is covered by a sieve. The reservoir is shaken, and particles leave it through the sieve. The amount of injected particles is not clearly defined, since it depends on unknown factors such as the initial location of particle in the reservoir, or the amount of clogging in the sieve holes. Therefore, the particle number can vary between experiment runs, since after each parabola, particles usually drop onto the lower electrode and are lost, and new particles have to be injected in the next run. 

The pulse sequence was employed during $13$ parabolas for different electrode separations $\Delta L$ ($45$, $75$~mm), neutral gas pressures $P$ ($5$, $20$~Pa), peak-to-peak voltages $V_{\rm{pp}}$ ($70$, $90$~V), and several base periods $T_{\rm{base}}$ in the range $50-800$~$\mu$s. The voltage $V_{\rm{pp}}$ is adjusted to the same value for each individual electrode segment. Since at each time only one electrode segment is on, $V_{\rm{pp}}$ is two times the voltage amplitude with respect to ground.
The chosen range of $T_{\rm{base}}$ ensures that the switching between channels happens at frequencies too large for the microparticles to follow: typical dust plasma frequencies for particles with a few micrometer diameter are in the range $35-1000$~Hz. 

The pulsed mode was either switched on before the start of the microgravity phase, or shortly after particle injection. In two parabolas $T_{\rm{base}}$ was changed during the parabola to see the direct effect on the particle cloud at a constant particle number. 
Therefore, $15$ experimental runs with different pulse settings were available for analysis.
The experimental parameters are summarized in table~\ref{table:experiments}. 

\section{\label{sec:results}Results}

The first qualitative observation in all performed experiments was that with starting the random pulse mode the central void was instantaneously closed, as shown in Fig.~\ref{fig:fig1}(b). This behaviour was always reproducible and did in general not depend on neutral gas pressure, rf power, electrode separation or pulse base period within the tested parameters.

To obtain a more quantitative picture of this observation, the changes in the area accessed by the particles (represented by the two-dimensional cross section of the cloud), and the particle density and distribution within this area (representing the homogeneity within the cloud) are investigated in more detail. This analysis is performed based on the individual particle coordinates, which are obtained from the recorded images such as Fig.~\ref{fig:fig1} by an intensity weighting method \cite{mohr2019}:

Connected groups of pixels with intensities above a user-chosen threshold are identified as particles, and the particle position is calculated as the intensity weighted center of the pixel group. To improve the results and compensate for possible nonuniformities in the illumination, the threshold was provided by local thresholding with a DOG-filter (difference-of-Gaussian).

\subsection{Cloud area and shape}\label{sec:shape}

\begin{figure}
  \includegraphics[width=\columnwidth]{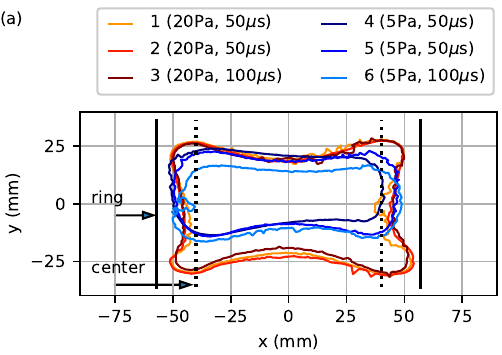}
  \includegraphics[width=\columnwidth]{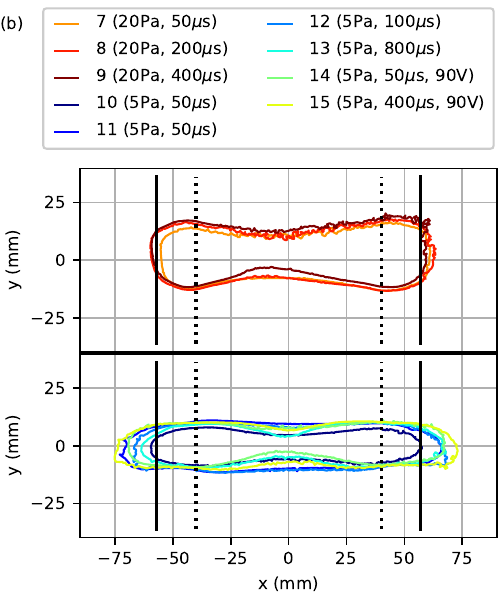}
 \caption{\label{fig:fig3}Cloud shapes obtained from the alpha shape method as explained in the main text, for electrode separations (a) $75$~mm, and (b) $45$~mm. The numbers in the legends are the data set numbers defined in table~\ref{table:experiments}. Neutral gas pressure and $T_{\rm{base}}$ are given in parentheses. $V_{\rm{pp}}$ was $70$~V for all experiments except \#14-15. For better visibility, the shapes in (b) are plotted separately for pressures $20$~Pa (top) and $5$~Pa (bottom). }
\end{figure}

The cloud shape can be extracted from the particle coordinates as the envelope of the visible cloud. In the two-dimensional cross section provided by the images, the shape represents the area accessed by the particles. It is determined by the balance of forces acting on and between the particles: the electric confinement force pointing from electrodes or grounded walls towards the bulk plasma for the negatively charged particles, and the outward pointing ion drag force and the mutual repulsion between the particles.

The shape of the electric confinement, and thus the cloud shape, depends on the chamber and electrode geometries, which determine the shape and location of the bulk plasma.
Since a four-electrode system is used, the plasma geometry can potentially differ from that in a system with two plane-parallel electrodes, especially if the operation is not symmetric as in the case of the random pulsed mode described above. 

Figures~\ref{fig:fig3}(a),(b) show examples of the cloud shapes of the pulsed complex plasma for two different electrode separations $45$ and $75$~mm. 
The envelopes are obtained from single images using the alpha shape method \cite{edelsbrunner1983}. This method can identify shapes which are not necessarily convex from data given as sets of points, in our case the particle coordinates.

\begin{figure}
\includegraphics[]{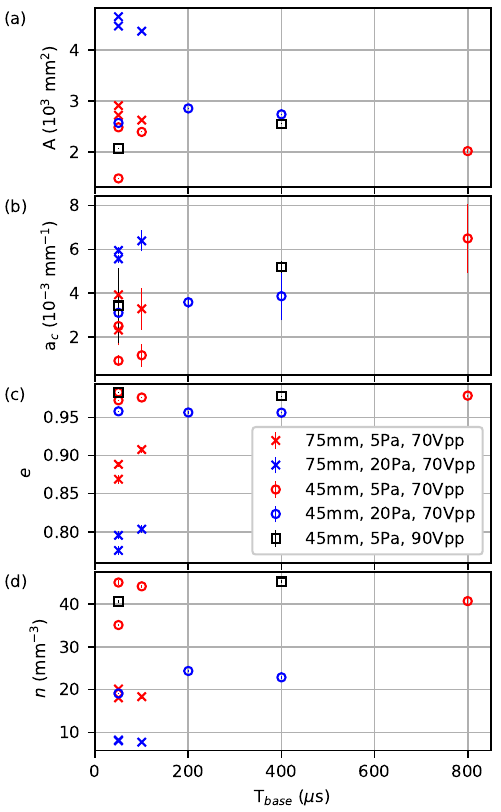}
\caption{\label{fig:fig4} (a) Area $A$ of the particle cloud, averaged over 100 frames. (b) Curvature $a_{c}$ of the horizontal cloud edges obtained as the time average of polynomial fits to 100 consecutive frames.
(c) Cloud eccentricity $e$ quantifying the elliptic shape of the cloud (in horizontal direction) with $e=0$ for a circle, $e$ growing towards $1$ for an increasingly elliptic shape. (d) Particle density $n$ from coordinates as described in the main text, averaged over 100 frames. All parameters are plotted vs. the pulse base period $T_{\rm{base}}$.
Error bars for (b) are derived from the uncertainty of the fit parameters, all other error bars (standard deviation of the mean) are of the size of the plot symbols.}
\end{figure}

The shape is not necessarily elliptical, and even can become concave with dents at the upper, lower, right and left sides (see e.g. Fig.~\ref{fig:fig3}(a) for $\Delta L=75$~mm and $P=20$~Pa.). At large electrode separation and higher pressure, the vertical cloud extent is larger than at low pressure, presumably due to the larger sheath extent at low pressure. The horizontal extent is not affected. At small electrode separation, the cloud expands horizontally for low pressure, while the vertical extent increases with increasing pressure, but the effect is less pronounced. The total cloud area, shown in Fig.~\ref{fig:fig4}(a), has no visible dependence on the pulse parameter $T_{\rm{base}}$, or the discharge voltage, and is considerably increased only at large $\Delta L$ and high pressure.
To quantify the cloud shape, the upper and lower horizontal edges (along $x$) of the envelopes of $100$ consecutive frames at times $t$ of each data set were fitted by the polynomial $a_{c}(t)x^{2} + b(t) x + c(t)$ with the fit parameter $a_{c}(t)$ describing the curvature of the edge. Figure~\ref{fig:fig4}(b) shows $a_{c}$ as average over all frames $t$, and upper and lower edges (chosing the positive sign). All data sets have a positive (negative) curvature at the upper (lower) edge (concave cloud shape), which is stronger pronounced for higher pressures at $\Delta L =75$~mm, and seems to increase with increasing $T_{\rm{base}}$ for $\Delta L =45$~mm.
In contrast, the shape in horizontal direction is concave only at  $\Delta L =75$~mm and $P=20$~Pa, and convex otherwise.

A possible reason for the more distinct concavity in y-direction could be the proximity of the electrodes: vertically, the particle cloud is confined in the space between the electrodes, and the electric field in the plasma sheath at the electrodes pushes the particles toward the chamber center. In horizontal direction, the grounded chamber walls are further away, allowing the cloud to expand. Also, the sheath width at the adjacent grounded guard rings should be smaller due to their area being larger than the driven electrode area by a factor of $1.65$, not including the chamber walls.

Another quantitative measure of the cloud shape is the eccentricity $e$. It can be used as a measure for how elliptical the shape is, not taking into account any irregularities such as dents or concave sections. It is defined as $e=\sqrt{(1-b_{\rm{min}}^2/b_{\rm{maj}}^2)}$ with $b_{\rm{maj}}, b_{\rm{min}}$ being the semi-major and semi-minor axes of the ellipse. For a circular cloud, $e$ is $0$, and it grows towards $1$ as the ellipse becomes more elongated in vertical direction (defining $b_{\rm{min}}$ as the semi-minor axis oriented in vertical direction).  The eccentricity was calculated from the eigenvalues of the covariance matrix of the second order central moments of the image moments \cite{flusser2016}, and is shown in Fig.~\ref{fig:fig4}(c). It can be seen that the cloud elongation in horizontal direction increases with decreasing pressure and decreasing electrode separation, but is does not strongly depend on $T_{\rm{base}}$. 

As becomes evident from the grouping of data points in Figs.~\ref{fig:fig4}(a)-(c), the two parameters affecting the cloud area and shape strongest are neutral gas pressure and electrode separation. The curvature (Fig.~\ref{fig:fig4}(b)) is the only parameter with an additional weak dependence on the pulse parameter $T_{\rm{base}}$. The limited amount of data for $\Delta L=75$~mm at larger $T_{\rm{base}}$ makes it difficult to draw a clear conclusion, and more experiments need to be performed and analyzed to fill that gap.

\subsection{Particle density}\label{section:particle_density}
The particle number density determines the dust plasma frequency. The presence of dust can also affect the plasma potential within the cloud, especially for large densities the plasma can be depleted of electrons that are then bound on the particles surface.
The density depends not only on the number of injected particles, which can vary between experiment runs due to the technical reasons described in section~\ref{sec:experiments}, but it is also influenced by the forces acting on the particles, since these determine the accessible area: injecting more particles can result in higher density, but also in cloud expansion if the repulsive force between the particles prevails. 

For each time step, the particle density is determined as the number of particles located inside the polygon calculated with the alpha shape method described in Sec.~\ref{sec:shape}, divided by the polygon area. This gives a time series of densities, that are then averaged over 100 frames to obtain an average density that can be compared between data sets.

Since the recorded images yield particle positions in two dimensions, i.e. representing the number of particles identified in the illuminated two-dimensional (2D) cross section viewed by the camera, the calculated density is a 2D particle density $n_{\rm{2D}}$.  To convert it to a 3D number density, $n_{\rm{2D}}$ is divided by the laser width $d_{l}$. The laser line shaping optics produce a line with width $0.16$~mm in the focus, and $0.22$~mm at the edge of the depth of focus, which is located approximately at the outer edge of the electrodes. 
An average value of $d_{l}=0.2$~mm is assumed. This approach is justified by the fact that the particle separations measured in the images are typically larger than the laser width ($370-885$~$\mu$m, depending on experimental conditions). Therefore, it is unlikely that more than one particle layer is visible, and the beam shape does not have an impact. The thus obtained density $n=n_{\rm{2D}}/d_{l}$ is a representative measure for the 3D particle density and its spatial variations, not the real number density which would require a different optical diagnostic system to measure three-dimensional particle positions in a larger volume.

Figure~\ref{fig:fig4}(d) shows  $n$ plotted vs. $T_{\rm{base}}$. As for the cloud area and eccentricity, no correlation with the base period or discharge voltage can be detected, but clearly $n$ decreases with increasing $\Delta L$ and pressure. 

\subsection{Density gradients}\label{section:density_gradients}

\begin{figure}
 \includegraphics[]{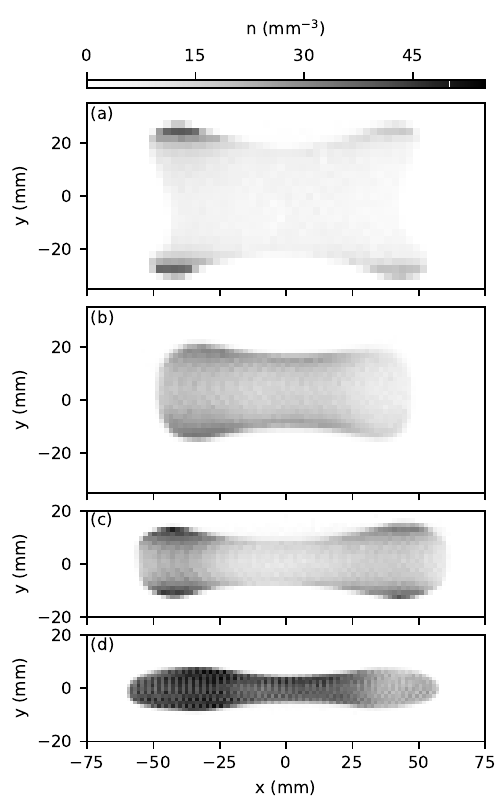}
  \caption{\label{fig:fig5}Density maps for data sets with $T_{\rm{base}}=50$~$\mu$s, $V_{\rm{pp}}=70$~V and (a) $\Delta L=75$~mm, $P=20$~Pa (data set \#1) (b) $\Delta L=75$~mm, $P=5$~Pa (data set \#5) (c) $\Delta L=45$~mm, $P=20$~Pa (data set \#7) and (d) $\Delta L=45$~mm, $P=5$~Pa (data set \#10).}
\end{figure}

\begin{figure*}
  \includegraphics[]{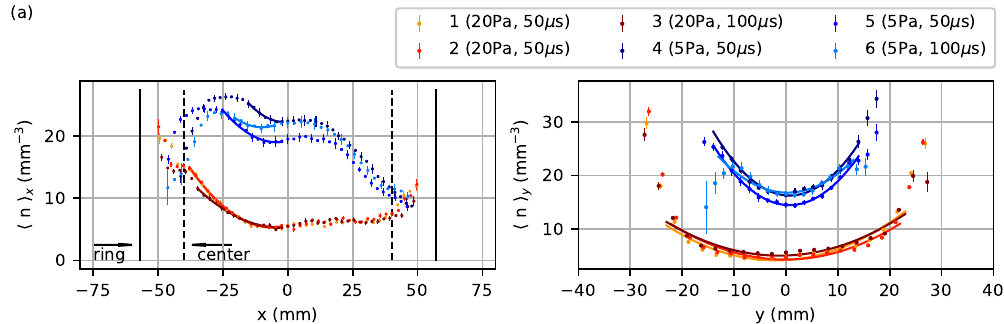}
  \includegraphics[]{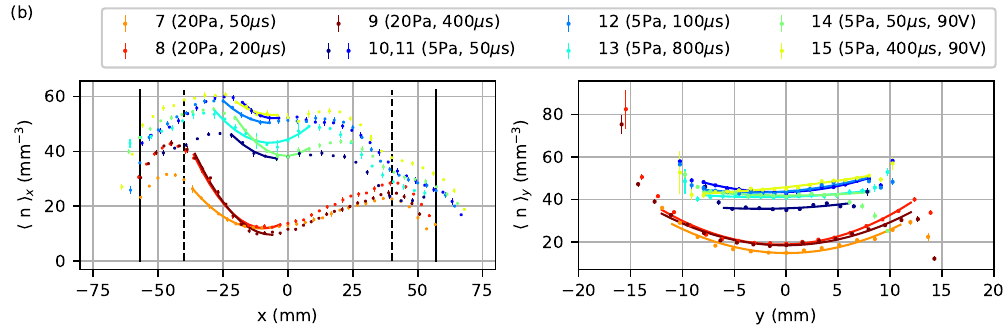}
\caption{\label{fig:fig6}Density profiles (colored dots) for (a) $\Delta L=75$~mm and (b) $\Delta L=45$~mm. The left (right) column shows the profile along $x$ ($y$).  Error bars are standard deviations from averaging over $100$ frames. The numbers in the legends are the data set numbers defined in table~\ref{table:experiments}, neutral gas pressure and pulse base period are additionally given in parentheses.
The position of the central and ring electrode outer edges are indicated by vertical solid and dashed black lines in the $x$ profiles (left images). The top and bottom electrode positions in the vertical profiles (right images) are at $y=\pm 37.5$~mm (a) and $y=\pm 22.5$~mm (b), respectively.
The fitted quadratic functions are shown as solid lines with colors corresponding to the data set color.}
\end{figure*}

The homogeneity of the density across the cloud is expected to be affected by the shape of the electric potential in the plasma chamber. Regions of higher potential attract the negatively charged particles, but from these regions positive ions are streaming outward.
The resulting ion drag force on the particles is pointing outwards, counteracting the confining electric forces. If the ion drag force is weak, there might be no visible central void, but still there could be spatial inhomogeneities in the particle density.

The density variations across the particle clouds can be illustrated by density maps obtained from the particle images: 
The image area is divided into a regular grid with a grid size of $3\Delta$, with $\Delta$ being the average particle separation in a given data set. The chosen grid size ensures that a grid cell in one image contains on average 10 particles.
The density of a cell is now calucated as the number of particles in the cell averaged over $100$ consecutive frames (keeping the spatial cell location unchanged), and divided by the cell area.
Figure~\ref{fig:fig5} shows examples of the density maps for selected electrode separations and pressures.

A qualitative observation is that at large $\Delta L=75$~mm, the density is in general higher in the four cloud corners compared to the cloud center. This is most pronounced at high pressure ($20$~Pa). At $5$~Pa, the density is on average higher across the whole cloud, and regions of higher density extend along the top and bottom cloud edges, as visible in Fig.~\ref{fig:fig5}(b).
A similar picture holds for the smaller electrode separation of $45$~mm (Fig.~\ref{fig:fig5}(c),(d)), though here the density increase at the cloud corners is accompanied by a general higher density in the outer (left and right) regions at $20$~Pa. 
At $5$~Pa, the cloud becomes very dense and the density becomes more uniform across the cloud.

In all examples, the density appears to be lower in the right image half. One factor contributing to this phenomenon is the illumination: the laser illuminating the particles is shining into the chamber from the left image side. At the right side some of the light is already scattered and the intensity in the laser sheet is therefore reduced, resulting in weaker illumination (thus a reduced particle detection rate and seemingly reduced $n$). Another factor could be a small residual acceleration (visible in the acceleration data of the flights) due to the flight maneuver towards the left in the images acting on the particles and increasing the density on that side. 

The spatial density gradients can be quantified by calculating horizontal and vertical density profiles: 
The horizontal density profile is obtained by dividing the particle cloud into vertical slices, or bins. The density in each slice is then obtained as the number of particles in the slice, divided by the slice area. The bin center on the horizontal $x$-axis is taken as the slice center. The width of the slices is chosen for each data set to be at least $3\Delta$ (the same as the cell grid size for the density maps). Since the vertical extent of the slices is across the full particle cloud height, which is much larger than the width of the slice, the number of particles in each slice is usually larger than $100$.
 The binning is repeated for the same $100$ frames used for the density map calulcation in Sec.~\ref{section:particle_density}. 
To account for temporal drifts of the cloud, for each frame the center position (the average over all particle positions) of the cloud is subtracted.
Finally, each slice's density is averaged over the $100$ frames. 

The same procedure is applied to obtain vertical density profiles, with slices taken along the horizontal direction.

Figure~\ref{fig:fig6}(a),(b) show the horizontal and vertical density profiles for $\Delta L = 75$~mm and $45$~mm, respectively. In the horizontal profiles (left images in Fig.~\ref{fig:fig6}(a),(b)), a decrease in density in the positive $x$-direction can be seen which is due to the lower particle detection rate in the right half of the images as explained above.

The horizontal profiles at $\Delta L =75$~mm (Fig.~\ref{fig:fig6}(a), left) at $P=5$~Pa exhibit a double peak structure with the two maxima located within the central electrode confines (data sets \#4-6), and a parabolic shape between the peaks. The minimum is located near the chamber center with a slight offset, presumably caused by residual forces during the flight maneuver.
At $20$~Pa (data sets \#1-3), the density is in general lower, and there are no peaks but an increase of density towards the electrode edge. The perceived flattness of the profiles in the central region especially for positive $x$ is misleading, since there the particle detection rate was low. The pronounced density increase towards the edges is caused by the high density in the cloud corners, visible as dark areas in Fig.~\ref{fig:fig5}(a), which considerably increase the density of the outermost slices.
The vertical profiles (Fig.~\ref{fig:fig6}(a), right) exhibit a parabolic shape in the central region, indicating a parabolic potential gradient between the electrodes positioned at $y=\pm37.5$~mm. This indicates that regardless of the random pulsing of the four channels and their geometric configuration, the bulk plasma is -- on time average -- located in the central volume of the chamber. As for the horizontal profiles, the vertical profiles are flatter at higher pressure (omitting the outer edges where the statistic is expected to be poor), while the overall density is larger for the lower pressure.

At $\Delta L =45$~mm (Fig.~\ref{fig:fig6}(b)), the double peak structure in the horizontal profiles (left image) appears also for the high pressure case (data sets \#7-9), with the two maxima located within or at the edge of the central electrode. Again the density is higher for $5$~Pa (data sets \#11-15).
The vertical profiles (right image) are parabolic at $20$~Pa with a lower density, and they become comparably flat at low pressure.  
The electrodes are located at $y=\pm22.5$~mm in the right image.

Comparing the two sets at different $\Delta L$, it is obvious that the density for the smaller $\Delta L$ is approximately twice as large along the profiles, in agreement with the averaged density $n$ (Fig.~\ref{fig:fig4}(d)).

Horizontal ($x$)  and vertical ($y$) profiles were fitted by quadratic functions $a_{\xi}\xi^{2} + b_{\xi}\xi+c_{\xi}$  with $\xi=x$, $y$, respectively. The fit parameters $a_{\xi}$, $b_{\xi}$ and $c_{\xi}$ describe the curvature, the displacement and the vertical offset. 
The fits are shown as solid lines in Fig.~\ref{fig:fig6}. Due to the decrease in density in the positive x-direction the fit to the horizontal profiles was performed only in the left part of the profiles, where the parabolic shape was still clearly visible. The left cut-off was chosen either as the position of the maximum density value, or at the edge of the center electrode indicated by the dashed line if no local maximum was detectable.

For the vertical profiles the data range for the fit was chosen such that the best agreement was in the central region, omitting the edges if necessary. The homogeneity in the central region, where edge effects have less influence on the particle dynamics, is in general of larger interest for scientific studies.

The fit parameters $a_{x}$, $a_{y}$ as a measure for the curvatures are presented in Fig.~\ref{fig:fig6} plotted vs. $T_{\rm{base}}$, yielding a quantitative comparison of the homogeneity of the clouds at different experimental parameters. 

\begin{figure}
  \includegraphics[]{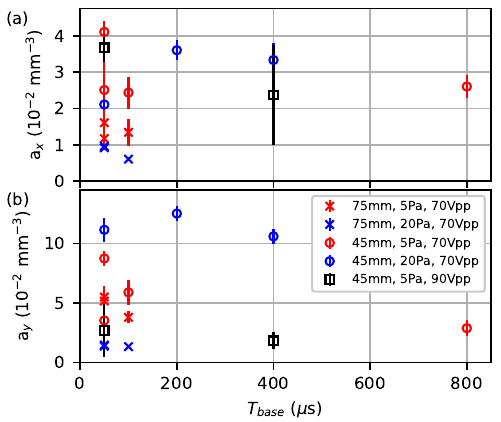}
  \includegraphics[]{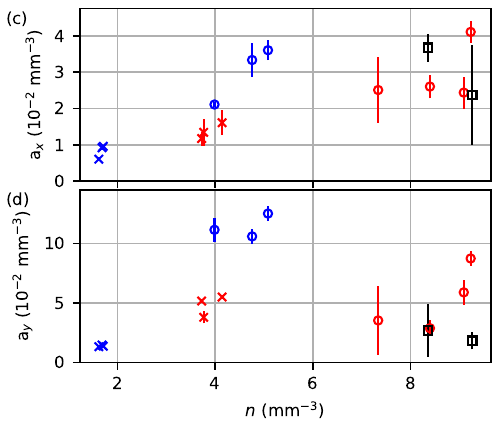}
\caption{\label{fig:fig7} Fit parameters $a_x$ (a) and $a_y$ (b) vs. pulse base period $T_{\rm{base}}$. The parameters represent the curvature of the density profiles in horizontal and vertical direction, and were obtained by a quadratic fits to the density profiles as described in the main text. (c), (d) show  $a_x$ and $a_y$ vs. particle density $n$.}
\end{figure}

The curvature of the horizontal profiles (Fig.~\ref{fig:fig6}(a)) is generally larger for $\Delta L=45$~mm, with no clear correlation to gas pressure, discharge voltage, or $T_{\rm{base}}$. For $\Delta L=75$~mm, the curvature is slightly larger at lower pressures. The vertical curvature (Fig.~\ref{fig:fig6}(b)) exhibits similar behavior, though here the curvature for $\Delta L=45$~mm increases with pressure, while it decreases with $P$ for $\Delta L=75$~mm. A distinct trend of either horizontal or vertical curvature with $T_{\rm{base}}$ can not be detected. 
As already mentioned before, more experimental data at different pulse periods are needed, and corresponding experiments are already scheduled, especially for the large electrode separation.

Figures~\ref{fig:fig6}(c),(d) show the horizontal and vertical curvatures vs. the particle density $n$. It becomes evident that $a_x$ and $a_y$ are grouped at specific $n$ according to neutral gas pressure and electrode separation, with no dependence on the discharge voltage $V_{\rm{pp}}$.

\section{\label{sec:discussion}Discussion}
The presented results indicate that the parameters with the largest influence on the particle cloud are neutral gas pressure $P$ and electrode separation $\Delta L$, with a weak influence of pulse base period $T_{\rm{base}}$ on the cloud shape. The observed dependencies and possible explanations are discussed in the following.

From the particle and energy balance for electropositive plasmas follows that increasing the neutral gas pressure decreases the electron temperature $T_{e}$, and increases the plasma density $n_{0}$ \cite{lieberman2005}. 

The effect of changing the electrode separation is more difficult to interpret. According to the self-consistent equations for the inhomogeneous discharge model for a cylindrical capacitively coupled rf discharge with plane-parallel electrodes \cite{lieberman2005}, if the sheath voltage is held constant, the plasma density can increase if the electrode separation is increased. The reason lies in the increase of the electric potential across the discharge in axial direction for increased discharge length (implying a larger power input into the plasma). This also affects the electron temperature (though not as pronounced as the influence of gas pressure): $T_{e}$ decreases with increasing electrode separation. 
It still needs to be investigated if this model is applicable to the geometrical configuration of the four-electrode system of the Zyflex chamber, and the (asymmetric) operation of the four electrode segments in the described pulsed mode, but it can at least give an indication of the expected influence on the plasma parameters.

In summary, $T_{e}$ decreases with increasing $P$ and $\Delta L$, $n_{0}$ increases with $P$ and $\Delta L$. 

That implies that, for large  $\Delta L$ and $P$ (low $T_{e}$, high $n_0$) the particle charges $Q_{p}$ become lower (since they depend on the mobility of electrons) and at the same time more ions are available that can induce a drag force on the particles. While the directed part of the ion drag force scales linearily with the ion density, the Coulomb part scales linearily with the ion density, and with the squared particle potential $Q_{p}^{2}$. The two effects (decreased $Q_{p}$, increased ion flow) can cancel out if the parameters are suitable and all other effects are ignored (e.g. change of particle charging due to streaming ions, change of ion velocity with pressure), and are therefore difficult to interpret without further data, e.g. from plasma simulations.

The pressure dependance of $T_{e}$ is well established and yields an decrease of the Debye length and sheath width with increasing $P$.
The bulk plasma volume increases, and the visible particle cloud area $A$ (representing the accessible volume as its 2D projection) increases. The effect is distinct for $\Delta L=75$~mm, but less pronounced for $\Delta L=45$~mm (Fig.~\ref{fig:fig4}(a)). Here, the larger areas at $\Delta L=75$~mm are due to the generally increased plasma volume.

The concave cloud shape (Fig.~\ref{fig:fig4}(b)) implies an inward pointing force from the electrodes toward the plasma center. A possible reason is the effect of the pulsed operation of the four channels on the ion flow:
on a time average, the ion flow direction could be manipulated such that it is pointing not from the chamber center outwards, but from a region close to the central electrode towards the adjacent ring electrode with both radial and axial components (on top and bottom chamber sides, respectively).
The ion drag force would then push the particles away from the central electrodes with a force component pointing towards the chamber center.
This could explain not only the absence of the central void, but also a stronger curvature at higher pressures and $\Delta L$ if the ion drag force increases at these conditions. The latter implies an increase of the ion flow $n_0u_i$ with the ion drift velocity $u_i$ (e.g. higher $n_0$, larger ion flow, stronger ion drag force). This assumption has to be examined further with the help of plasma simulations in a next step.
The increase of the curvature with increasing $T_{\rm{base}}$ might also indicate a further increase of the average ion drag force from the electrodes towards the bulk, caused by the longer time the directed ion flow acts on the particle cloud, as the frequency of switching between the channels comes closer to the dust plasma frequency ($1.25$~kHz at $T_{\rm{base}}=800$~$\mu$s).

The eccentricity $e$ (Fig.~\ref{fig:fig4}(c)) is larger for lower $P$, meaning an increasingly elliptic shape. Here, the sheath width becomes larger and the particle cloud is presumably more compressed in vertical direction between the electrodes, yielding an increase in the horizontal direction (i.e. the cloud is squeezed between the electrode sheaths). The measurements at lower $\Delta L$ yield a larger $e$, which could be due to the additional dependance of $T_e$ on $\Delta L$ (increase of $T_e$, increase of sheath width, more compression).

The increase of the particle density $n$ (Fig.~\ref{fig:fig4}(d)) with decreasing $P$ can also be attributed to the stronger compression between the sheaths. The decrease of $n$ for larger $\Delta L$ is contrary to the above arguments, though. An explanation could be as follows:
Assuming that the injected number of particles is at least approximately of the same order of magnitude (the number of shakes of the particle dispenser was the same), the density is expected to decrease in that cases since fewer particles are available to fill a larger area, respectively volume, at large $\Delta L$. 
The interpretation of the dependence of $n$ on the experimental parameters is therefore rather speculative without further measurements, especially without the ability to control the amount of injected particles more precisely. 
 Figure~\ref{fig:fig4}(d) gives a qualitative comparison of the experiment runs with regard to the particle density, though.

The curvatures of the density profiles can best be interpreted by the results presented in Fig.~\ref{fig:fig6}(c),(d).
The dependences of $a_x$ and $a_y$ on $P$ and $\Delta L$ are in agreement with the dependence of $n$ on $P$ and $\Delta L$ (Fig.~\ref{fig:fig4}(d)). The increase of $a_{y}$ with $P$ at $\Delta L=45$~mm is further in agreement with the observation that the curvature of the cloud envelope in horizontal direction, $a_c$, increases with increasing pressure (Fig.~\ref{fig:fig4}(b)): the increase of $a_{c}$ with pressure indicates a stronger cloud compression in vertical direction, especially in the (horizontally) central cloud region, which in turn causes a stronger density gradient.

The reversion of pressure dependence at large electrode separation can be explained by the same argument as presented above for the density, that at the largest $\Delta L$ not enough particles were injected to fill the available volume and thus the results might be misleading.

The vertical density gradients are generally larger than the horizontal gradients ($a_y$ is nearly one magnitude larger than $a_x$). Again this can be explained by the stronger potential gradients and thus confinement forces between the driven electrodes, compared to the confinement in horizontal direction that mainly depends on the potential difference between plasma and grounded surfaces (guard ring, chamber walls).

\section{\label{sec:conclusion}Conclusion and Outlook}

A new technique for closing the, often unwanted, central void in a cloud of micrometer-sized particles immersed in the bulk of a low-temperature plasma in microgravity has been presented.
The technique involves the utilization of a four-channel rf generator, connected to four separate electrodes in the plasma chamber, that is operated in a pulsed mode: each of the channels is pulsed by applying a random sequence of on- and off-times. The sequences are chosen such that at each time period $T_{\rm{base}}$ only one channel (one electrode) is in the ``on'' state (random pulsing). The random sequence (given in the Appendix) was chosen to omit potential effects due to short-term periodicity.

First observations of $15$ experimental runs performed during a parabolic flight campaign at different discharge conditions show that void closure appears independent on the parameters neutral gas pressure, discharge power, or discharge volume, as long as the rf generator was operated in the pulsed mode.

The characteristics of the particle system for different experimental parameters was investigated qualitatively and quantitatively by analyzing properties of the shape of the particle cloud (area, edge curvature and eccentricity), and the particle density and density gradients.
While the pulse sequence was not changed in this first set of experiments, the pulse base period $T_{\rm{base}}$ was modified to investigate the influence of this pulse parameter on the particles. The results show that there is no distinct indication that the investigated particle cloud characteristics depend on $T_{\rm{base}}$ within the tested parameter range. A weak dependence of the cloud curvature on $T_{\rm{base}}$ could indicate an influence on the magnitude of the ion flow and thus the ion drag force. This observation requires further investigations of the underlying plasma processes, which will also help to uncover the reasons for void-closure that are presumed to be connected to the direction of the ion flow in the special conditions of the pulsed operation mode.

It was found that neutral gas pressure and electrode separation (plasma volume, or discharge geometry) have the strongest influence, with larger electrode separation yielding larger and more homogeneous particle clouds, while larger pressures seem to cause larger density gradients in vertical direction - the later being an observation that needs more experimental data to be confirmed.

The minor influence of $T_{\rm{base}}$ is an interesting result since it implies that the pulsed mode does not introduce new physical effects on the particles aside from the void closure, thus allowing to perform studies of fundamental processes in extended dust systems without the need to adapt the existing theoretical models to include specific effects of the pulsed mode.

Even though the density of the particle cloud, as analyzed in section~\ref{section:density_gradients}, is not perfectly uniform and shows gradients depending on gas pressure and electrode separation, the void-free system provides access to large (3D) particle clouds without inner boundaries. Tuning of the experimental conditions could further improve the homogeneity, e.g. using larger electrode separations.

The proposed technique shows great potential for application in experiments dedicated to fundamental studies of complex plasma, for example in the future complex plasma facility COMPACT \cite{knapek2022} that is currently under development, and is intended to be operated onboard an orbital platform, e.g the International Space Station (ISS).

As evident from the presented data, a larger amount of data sets is needed to expand the parameter space and get a deeper understanding of the underlying processes, especially the physical reason for void closure. 
A detailed experimental program, to be conducted on parabolic flights, has been worked out to investigate the influence of e.g. periodic sequences, the necessity of a four electrode system, and a wider range of base periods at different gas pressures, discharge voltages and volumes. In addition, plasma simulations are needed to shed light on the modification of plasma parameters during the pulsed operation. The simulations can yield plasma density and electron temperature, but also the spatial ion flow pattern and electric fields in the chamber that will give further insight into the mechanism of the void closure.

\begin{acknowledgments}
  This work and the authors were funded by DLR/BMWi (FKZ 50WP0700, FKZ 50WM1441) and the Bavarian Ministry of Economic Affairs and Media, Energy and Technology (StMWi). C. A. Knapek and D. P. Mohr are currently funded by DLR/BMWi FKZ 50WM2161. We thank H.~M.~Thomas for carefully reading the manuscript.

\end{acknowledgments}

\appendix
\section{\label{sec:appendixA}Pseudorandom Sequence}

The sequence was generated with Python and the numpy package in the following way: at first, a list of length $1024$ was generated, with equal occurence of the numbers {0, 1, 2, 3}, i.e. 256 occurences of each element. The numbers represent the channels of the rf generator, connected to the electrode segments, as indicated in Fig.~\ref{fig:fig2}: top center (0), top ring (1), bottom center (2), bottom ring (3).

Then, the list was shuffled using the function numpy.random.shuffle, which is based on the Mersenne Twister \cite{matsumoto1998}. The result was a list of pseudorandomly distributed numbers {0, 1, 2, 3}. During the presented experiments, accidentally only the first $256$ entries of the list were used:
 
\seqsplit{
  0113001021302210103301132010311303010301232202101100021212211221011223222333023302311221103032102200021113132223122112123100331013003300211022332122221221112130223233122121012133300223020131312102010203101201320220212323210100221000301013312032103330330200}

Each number in the list represents a time interval of the length of the base period $T_{\rm{base}}$. The respective number determines which channel is in the ``on''-state at this time interval, i.e. 0 means channel 0 is on.
The list is converted into a sequence of Bytes, each Byte containing the state of all four channels for one specific time step, with a true Bit describing the on-state of a channel.
This sequence was loaded into the memory of an Arduino Nano microcontroller which triggered the four channels of the rf generator via four digital signal outputs (high or low for on- or off-states, respectively) on one register (resulting in simultaneous switching), with the base period $T_{\rm{base}}$ as the free parameter.

\begin{figure}[h]
  \includegraphics[]{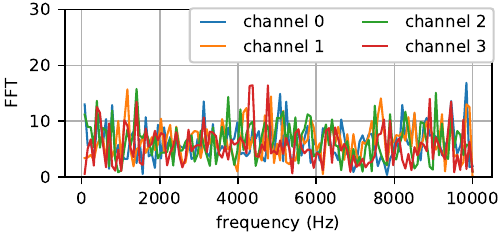}
\caption{\label{fig:fig_appendix} Fast Fourier transform of each channel's occurence (list of on- and off-states for each channel).}
\end{figure}

To illustrate the uniformity of the sequences, Fig.~\ref{fig:fig_appendix} shows the frequency spectra obtained from a Fast Fourier transform of each channel's occurence in the shortened list of length $256$, as a basic statistical test for a random sequence exhibiting pseudorandom noise over all detectable frequencies. 

\bibliography{knapek_pulsed_complex_plasma}

\end{document}